\documentclass[lettersize,journal]{IEEEtran}
\usepackage{amsmath,amsfonts}
\usepackage{amssymb}
\usepackage{algorithmic}
\usepackage{algorithm}
\usepackage{array}
\usepackage[caption=false,font=normalsize,labelfont=sf,textfont=sf]{subfig}
\usepackage{textcomp}
\usepackage{stfloats}
\usepackage{url}
\usepackage{verbatim}
\usepackage{graphicx}
\usepackage{cite}
\usepackage{booktabs}
\newcommand{\eq}[1]{Eq. \eqref{#1}}
\begin{document}

\title{CPMamba: Selective State Space Models for MIMO Channel Prediction in High-Mobility Environments}

\author{Sheng Luo,~\IEEEmembership{Member,~IEEE,}
        Jiashu Xie, Yueling~Che, \IEEEmembership{Member,~IEEE}, Junmei~Yao, Jian~Tian, \IEEEmembership{Member,~IEEE}, Daquan~Feng, \IEEEmembership{Member,~IEEE}, and Kaishun~Wu, ~\IEEEmembership{Fellow,~IEEE}
\thanks{Sheng Luo, Jiashu Xie, Yueling Che, Junmei Yao are with the
College of Computer Science and Software Engineering, Shenzhen University,
Shenzhen 518060, China (e-mail: sluo@szu.edu.cn; 2410103001@mails.szu.edu.cn; yuelingche@szu.edu.cn; yaojunmei@szu.edu.cn).}

\thanks{Daquan Feng is with the College of Electronics and Information Engineering, Shenzhen University, Shenzhen, China (e-mail: fdquan@szu.edu.cn).}

\thanks{Jian Tian is with TikTok Inc., Culver City, CA 90230, USA (e-mail: tianjian.colin@tiktok.com).}

\thanks{Kaishun Wu is with the Hong Kong University of Science and Technology, Guangzhou 511458, China, and also with the College of Computer Science and
Software Engineering, Shenzhen University, Shenzhen 518060, China (e-mail: wu@szu.edu.cn).}
}



\maketitle

\begin{abstract}
Channel prediction is a key technology for improving the performance of various functions such as precoding, adaptive modulation, and resource allocation in MIMO-OFDM systems. Especially in high-mobility scenarios with fast time-varying channels, it is crucial for resisting channel aging and ensuring communication quality. However, existing methods suffer from high complexity and the inability to accurately model the temporal variations of channels. To address this issue, this paper proposes CPMamba—an efficient channel prediction framework based on the selective state space model. The proposed CPMamba architecture extracts features from historical channel state information (CSI) using a specifically designed feature extraction and embedding network and employs stacked residual Mamba modules for temporal modeling. By leveraging an input-dependent selective mechanism to dynamically adjust state transitions, it can effectively capture the long-range dependencies between the CSIs while maintaining a linear computational complexity. Simulation results under the 3GPP standard channel model demonstrate that CPMamba achieves state‑of‑the‑art prediction accuracy across all scenarios, along with superior generalization and robustness. Compared to existing baseline models, CPMamba reduces the number of parameters by approximately 50\% while achieving comparable or better performance, thereby significantly lowering the barrier for practical deployment.
\end{abstract}


\section{Introduction}
\IEEEPARstart{M}{ultiple-Input} Multiple-Output (MIMO) technology serves as a fundamental enabler for improving spectral efficiency in fifth-generation (5G) and next-generation wireless communication systems\cite{mimo}.
The performance of key physical-layer functions---including beamforming\cite{beamforming}, adaptive modulation\cite{adaptive}, and resource allocation\cite{resource}---heavily depends on the availability of real-time and high-precision channel state information (CSI). However, in high-mobility scenarios (such as vehicle-to-everything and UAV communications) \cite{uav_beam}, the Doppler effect induces rapid channel time-variation that severely compress coherence time and lead to significant channel aging. These effects considerably degrade the throughput and transmission reliability. As a result, accurate prediction of future channel responses based on historical CSI has become an essential technique for maintaining communication efficiency under such dynamic conditions.

The core challenge in channel prediction is to accurately capture the temporal dynamics of CSI. Traditional model-driven approaches---including parametric models \cite{param_ref}, Kalman filtering \cite{Kalman_filtering}, autoregressive (AR) modeling \cite{ar_ref}, and linear extrapolation \cite{linear_extrapolation}---rely on explicit statistical or physical assumptions about the channel. Although these methods are computationally efficient, their reliance on theoretical models limits their ability to capture realistic multipath characteristics. As a result, prediction accuracy degrades significantly in complex propagation environments featuring nonlinearities and rich scattering.

Deep Learning (DL), as a leading branch of artificial intelligence (AI), has revolutionized numerous fields—including healthcare \cite{medical}, autonomous driving \cite{driving}, recommender systems \cite{recommender}, and financial services \cite{stock}—and has also opened up new possibilities for channel prediction. Recurrent Neural Networks such as RNN \cite{rnn_survey_ref} and LSTM \cite{lstm} capture short-term dependencies through temporal inductive bias. However, their sequential computation limits parallelization efficiency, and gradient vanishing issues\cite{hochreiter2001gradient} hinder their ability to model long-term dynamics. Other architectures, such as convolutional neural networks (CNNs) \cite{cnn_cp,cnn2}, along with generative adversarial networks (GANs), have been applied to channel prediction by treating channel data as images for temporal feature extraction, enabling parallel prediction of multiple future CSI frames. Nevertheless, these approaches often fail to fully account for the inherent structural properties of wireless channels.

The Transformer architecture \cite{transformer} has demonstrated breakthrough performance in channel prediction by leveraging the powerful modeling capabilities of its self-attention mechanism. Related pre-trained large language models (LLMs) have also been introduced for channel prediction tasks \cite{llm4cp, wifo}. However, their high computational complexity results in quadratic growth in memory consumption and inference latency when processing high-dimensional CSI data, which poses significant challenges for deployment on resource-constrained edge base stations.

Recently, the Mamba architecture \cite{mamba} introduces selective state space models (SSMs) that reduce the complexity of sequence modeling to linear $O(N)$. This framework achieves efficient computation while preserving a global receptive field, offering novel insights for channel prediction. Its selective mechanism dynamically adapts state transitions according to the input, theoretically aligning with the non-stationary Markov evolution of time-varying channels. Meanwhile, the linear complexity provides intrinsic advantages in processing long CSI sequences. As the first method that applies Mamba to channel prediction, ChannelMamba \cite{channelmamba} effectively models temporal dependencies through dual-domain input embedding and cross-path parameter-shared Mamba modules, achieving efficient CSI temporal prediction.

In this paper, we adopt the Mamba architecture as the core module for channel prediction, capitalizing on its efficient sequence modeling capability to address channel prediction challenges across diverse scenarios. Specifically, we propose CPMamba, an end-to-end deep learning model, with the following key contributions:
\begin{itemize}
\item We propose a Mamba-based channel prediction method for MIMO-OFDM systems in high-mobility environments. This method leverages the efficient sequence modeling capability of the Mamba architecture to achieve effective channel prediction while maintaining low computational complexity.
\item In the proposed channel prediction framework, we have designed a dedicated feature extraction and embedding module for channel data, which effectively transforms and embeds the channel data into feature space. For temporal prediction, we developed a versatile Mamba module that, through residual connections and serial stacking, enables efficient temporal modeling of channel variations.
\item Extensive experimental evaluations demonstrate that the proposed CPMamba model achieves state-of-the-art (SOTA) performance in channel prediction tasks, surpassing existing baseline methods across various standard simulation scenarios.
\end{itemize}

This paper is structured as follows: Section \ref{related_work} reviews related work. Section \ref{system_model} establishes the system model and formulates the problem. Section \ref{preliminary} introduces the fundamentals of SSMs and preliminary knowledge of the Mamba architecture. Section \ref{method} provides a detailed description of the proposed CPMamba framework. Section \ref{experiment} presents and discusses the experimental results. Finally, Section \ref{conclusion} concludes the paper.

\section{Related Work}
\label{related_work}
This section reviews relevant literature on channel prediction in two parts: first, a survey of channel prediction methods, categorized into traditional model-based and data-driven deep learning approaches; second, a focus on recent advances in the Mamba architecture, a core component of our proposed framework.

\subsection{Traditional Channel Prediction}

Traditional channel prediction research mainly relies on parametric or linear prediction models. For example, reference \cite{param_ref} estimates multipath parameters like angle of arrival and Doppler shift from received signals, and extrapolates future channels under short-term stationarity assumptions. Reference \cite{ar_ref} employs AR models to characterize flat fading channels, determining the model order and parameters based on channel statistics such as autocorrelation functions. In reference \cite{linear_extrapolation}, the current phase and its rate of change (frequency) for each multipath component are estimated from historical channel measurements; the phase values are then linearly extrapolated to future time instants to reconstruct the full channel response. Additionally, reference \cite{Kalman_filtering} adopts a data-driven Kalman filtering framework as a baseline, with experimental results demonstrating that advanced machine learning models can surpass traditional techniques in prediction accuracy.

While computationally efficient, these methods rely heavily on prior assumptions regarding channel behavior—particularly the premise of short-term channel stability. As a result, prediction performance degrades significantly in non-linear and rich-scattering propagation environments.

\subsection{Deep Learning-Based Channel Prediction}

Data-driven deep learning has yielded extensive results in channel prediction. The work of \cite{rnn_survey_ref} applies RNNs to construct frequency‑domain predictors for wideband communication and integrates them into MIMO-OFDM systems to enhance antenna selection accuracy. Reference \cite{lstm} incorporates LSTM or GRU units into a deep recurrent network for MIMO channel prediction, strengthening time‑series modeling through long short-term memory or gated memory mechanisms. However, such RNN‑based models suffer from low serial computational efficiency and gradient vanishing issues.

Another direction combines GANs with CNNs, treating channel state information as two‑dimensional images. The work of \cite{cnn_cp} combines a GAN variant based on mean squared error together with CNNs to infer downlink CSI from uplink observations. Reference \cite{cnn2} introduces a conditional GAN (CPcGAN) that uses a discriminator to measure distribution divergence between predicted and actual downlink CSI under uplink constraints, and a generator to learn the mapping from conditions to predictions. Such image‑based approaches, however, often overlook inherent channel structures.

The self‑attention mechanism of transformer architecture enables parallel computation across all sequence positions, offering powerful sequence modeling for channel prediction. Reference \cite{transformer} proposes a transformer-based parallel channel prediction scheme, formulating channel prediction as a
parallel channel mapping problem to predict channels in upcoming frames in parallel. Despite these breakthroughs, transformers suffer from quadratic complexity growth in memory and latency when processing high‑dimensional CSI, hindering practical deployment. Closely related LLMs have also been adapted for channel prediction \cite{llm4cp, wifo}, yet their substantial parameter scales pose challenges for communication‑scenario adaptation.

\subsection{Mamba-Based Models}

In recent years, SSMs\cite{gu2021combining,s4,hungry} have emerged as powerful tools for sequence modeling, garnering growing research interest. Among them, Mamba \cite{mamba}---as a selective SSMs---distinguishes itself through its input-dependent selective mechanism and hardware-aware parallel computing, delivering strong efficiency and performance on long sequences.

Mamba has achieved notable results in computer vision. By designing specialized scanning strategies for data, it can be applied to visual \cite{vmamba} and other non‑sequential domains. Specifically, reference \cite{vmamba} introduces a cross‑scanning module to convert image data into sequences, retaining benefits from CNNs and Vision Transformers (ViT) while improving computational efficiency without sacrificing the global receptive field. Reference \cite{mambavision} proposes a hybrid Mamba‑Transformer backbone that reformulates Mamba's structure to enhance its capacity for visual feature modeling.

In time‑series prediction, Mamba has also drawn considerable attention due to its linear complexity and strong long‑sequence modeling capability, with studies confirming its effectiveness \cite{mamba_effective}. Bi‑Mamba+ \cite{bimamba_plus} augments Mamba with a forget gate to better preserve historical information, selectively merging new and past features in a complementary manner. CMMamba \cite{cmmamba} employs bidirectional Mamba to model sequences and uses a channel mixing mechanism to enrich feature representation within original channels. TimeMachine \cite{timemachine} generates multi‑scale contextual cues and adopts a novel quadruple Mamba architecture to unify channel‑mixed and channel‑independent modeling, handling different temporal scales separately to integrate long‑term and short‑term patterns.

In summary, the Mamba architecture offers an efficient and powerful sequence model with linear computational complexity and good scalability, showing strong potential across multiple fields and providing a theoretical foundation for channel prediction. To fully adapt it to channel prediction, however, tailored sequence‑feature extraction modules must be designed according to the nature of channel data.

\section{System Model And Problem Formulation}
\label{system_model}

\begin{figure}[t]
\center{\includegraphics[width=7.5cm]  {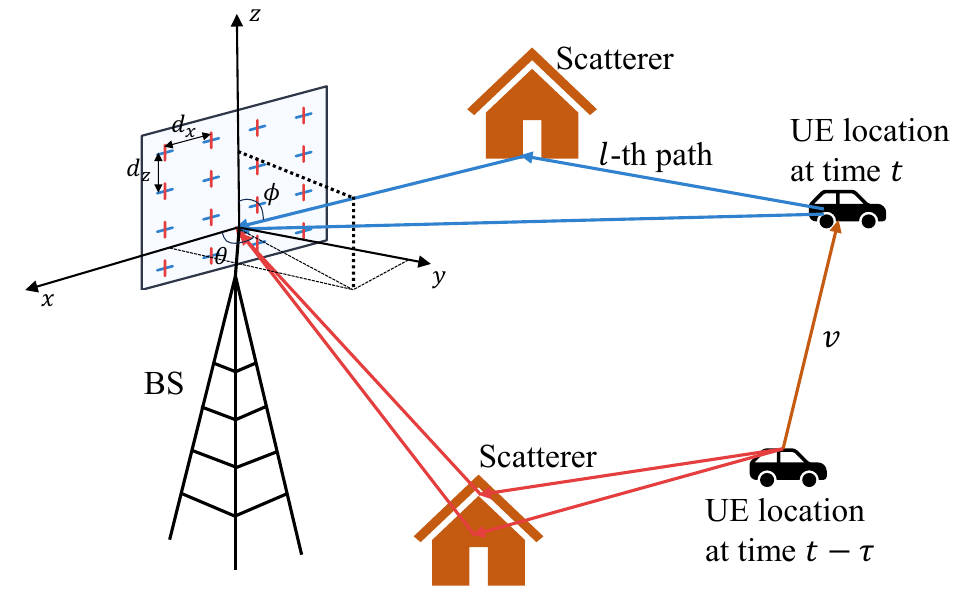}}
\caption{Geometric illustration of the multipath propagation environment involving a UPA-equipped BS and a high-mobility UE.}
 \label{fig_csi_est}
\end{figure}

\subsection{Channel Representation}
As shown in Fig.~\ref{fig_csi_est}, consider a massive MISO-OFDM downlink system in which the base station (BS) is equipped with a uniform planar array (UPA) comprising $N_t=N_h\times N_v$
transmit antennas, where $N_h$ and $N_v$ denote the number of antennas in the horizontal and vertical dimensions, respectively. The system operates over $K$ orthogonal subcarriers and can be extended to multi-user channel prediction through parallel processing of multiple samples. Assuming a channel model with $M$ discrete paths \cite{3gpp}, the channel state information for the $k$-th subcarrier at time $t$ can be expressed as:
\begin{equation}
\mathbf{h}_t[k] = \sum_{l=1}^{M} \alpha_t^l e^{j 2 \pi(v_l t- \tau_l f_k)} \, \mathbf{a}_t(\theta_l, \phi_l),
\end{equation}
where $\mathbf{h}_t[k] \in \mathbb{C}^{N_t \times 1}$ denotes the flattened channel vector aggregating the responses from both horizontal and vertical antenna dimensions, $M$ is the total number of resolvable multipath components, $\alpha_t^l$ represents the complex gain of the $l$-th path at time $t$, incorporating both amplitude attenuation and initial phase offset, $v_l$ and $\tau_l$ stand for the Doppler frequency shift and the propagation delay of the $l$-th path, respectively, while $f_k$ is the center frequency of the $k$-th subcarrier. Given a user equipment (UE) moving at speed $v$, if the angle between its velocity vector and the arrival direction of the $l$-th path is $\varphi_l$ , the corresponding Doppler shift can be expressed as $v_l = \frac{v}{\lambda} \cos \varphi_l$ where $\lambda = c / f_c$ is the signal wavelength, $c$ is the speed of light, and $f_c$ is the carrier frequency. The spatial steering vector $\mathbf{a}_t(\theta_l, \phi_l) \in \mathbb{C}^{N_t \times 1}$ captures the array response of the transmit UPA, with $\theta_l$ and $\phi_l$ being the azimuth and elevation angles of the $l$-th path, respectively.

For the UPA with respective antenna spacings $d_x$ and $d_z$ in the horizontal and vertical directions, the steering vector can be decomposed into horizontal and vertical components \cite{llm4cp}, given by:
\begin{equation}
\label{eqa_a}
\mathbf{a}_t(\theta_l, \phi_l) = \mathbf{a}_t^{h}(\theta_l, \phi_l) \otimes \mathbf{a}_t^{v}(\theta_l, \phi_l),
\end{equation}
where $\mathbf{a}_t^{h}(\theta_l, \phi_l)$ is the horizontal component and $\mathbf{a}_t^{v}(\theta_l, \phi_l)$ is the vertical component, given by:

\begin{subequations}
\label{eq:steering_vectors}
\begin{align}
\mathbf{a}_t^{h}(\theta_l, \phi_l) &= \bigl[1, e^{j\frac{2\pi}{\lambda} d_x\sin \phi_l\cos\theta_l}, e^{j\frac{2\pi}{\lambda} 2 d_x\sin \phi_l\cos\theta_l}, \cdots, \nonumber \\
&\qquad e^{j\frac{2\pi}{\lambda} (N_h-1) d_x\sin\phi_l\cos\theta_l}\bigr]^T \in \mathbb{C}^{N_h \times 1}, \\
\mathbf{a}_t^{v}(\theta_l, \phi_l) &= \bigl[1, e^{j\frac{2\pi}{\lambda} d_z\sin\phi_l \sin\theta_l}, e^{j\frac{2\pi}{\lambda} 2 d_z\sin\phi_l \sin\theta_l}, \cdots, \nonumber \\
&\qquad e^{j\frac{2\pi}{\lambda} (N_v-1) d_z\sin\phi_l \sin\theta_l}\bigr]^T \in \mathbb{C}^{N_v \times 1}.
\end{align}
\end{subequations}

Accordingly, the channel state information at time $t$, denoted as $\mathbf{H}_t$, can be expressed as follows:
\begin{equation}
\mathbf{H}_t=[\mathbf{h}_t[1],\mathbf{h}_t[2],\cdots \mathbf{h}_t[K]] \in \mathbb{C}^{N_t \times K}.
\end{equation}

Channel estimation based on general pilot signals is susceptible to noise \cite{channelmamba}. The acquired channel state information can thus be modeled as $\tilde{\mathbf{H}}_t=\mathbf{H}_t+\mathbf{Z}_t$, where $\mathbf{Z}_t\in\mathbb{C}^{N_t \times K}$ represents additive white Gaussian noise (AWGN) that distributed as $\mathcal{CN}(0,\sigma_w^2)$ with $\sigma_w^2$ denoting the noise power.

\subsection{Problem Formulation}
Channel prediction can be formulated as a time-series forecasting task: given a historical sequence of CSI over a window of length $L$, denoted as $\mathbf{S}_{p}=\{\mathbf{H}_{t-L+1},\mathbf{H}_{t-L+2},\dots,\mathbf{H}_{t}\} \in \mathbb{C}^{L \times N_t \times K}$, the goal is to predict the CSI for the subsequent $P$ time steps: $\mathbf{S}_{f}=\{\mathbf{H}_{t+1},\mathbf{H}_{t+2},\dots,\mathbf{H}_{t+P}\} \in \mathbb{C}^{P \times N_t \times K}$. Treating the predictor as a parametric mapping $f_\theta:\mathbb{C}^{L \times N_t \times K}\to \mathbb{C}^{P \times N_t \times K}$ and adopting the normalized mean square error (NMSE) as the performance metric \cite{transformer}, the problem can be formally expressed as the following constrained optimization:
\begin{subequations}
\begin{align}
\min_{\theta}\ &
\text{NMSE}(\hat{\mathbf{S}}_{f},\mathbf{S}_{f})=\frac{\lVert\hat{\mathbf{S}}_{f}-\mathbf{S}_{f}\rVert_{F}^{2}}
          {\lVert \mathbf{S}_{f}\rVert_{F}^{2}} \label{nmse_eqa}
\\
\text{s.t. } & \hat{\mathbf{S}}_{f}=f_{\theta}(\mathbf{S}_{p}),
\end{align}
\end{subequations}
where $\|\cdot\|_F^2$ represents the squared Frobenius norm, $\hat{\mathbf{S}}_f$ and $\mathbf{S}_f$ denote the predicted CSI and true future CSI, respectively, $\theta$ denotes the trainable parameters of the mapping $f_\theta$.

\begin{figure*}[ht]
\centering
\includegraphics[width=3.5in]{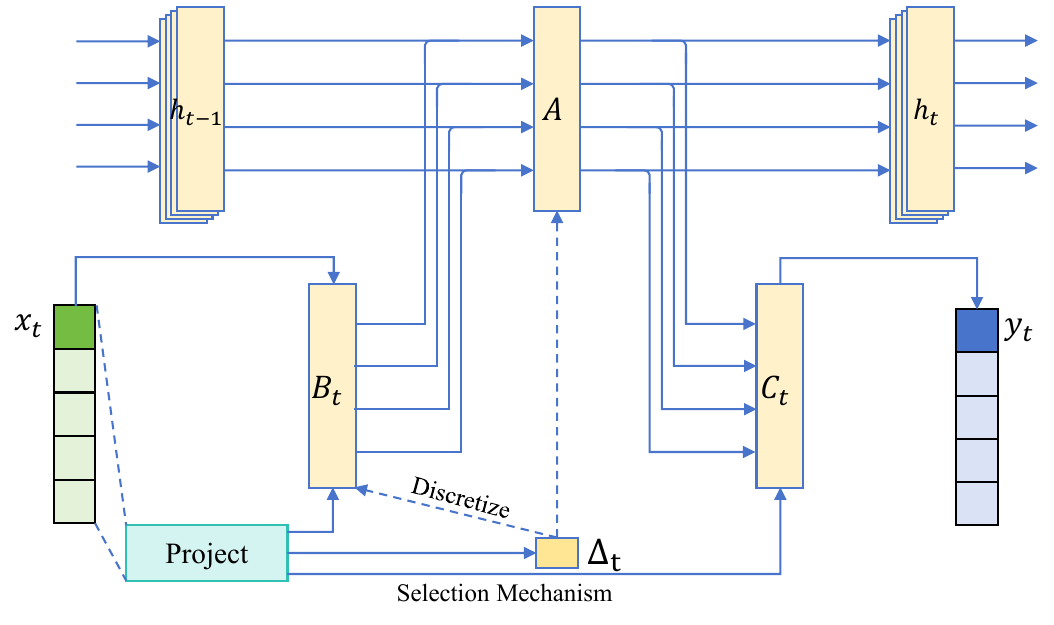}
\caption{The overall framework of SSM. }
\label{s6_fig}
\end{figure*}

\section{Preliminaries}
\label{preliminary}

\subsection{State Space Models}

The Structured State Space Sequence Model (S4)\cite{s4} is originally derived from a class of continuous-time dynamical systems\cite{pechlivanidou2022zero} that map a continuous-time input signal \(\mathbf{x}(t) \in \mathbb{R}^{D_{\text{in}}}\) to a continuous-time output \(\mathbf{y}(t) \in \mathbb{R}^{D_{\text{out}}}\) through an implicit latent state \(\mathbf{h}(t) \in \mathbb{R}^{N}\). Here, $D_{in}$, $D_{out}$ and $N$ represent the number of input signal dimension, output signal dimension, and latent state dimension, respectively. The system dynamics are governed by the following equation:
\begin{subequations}
\label{eqa_ss}
\begin{align}
\mathbf{h}'(t) &= \mathbf{A} \mathbf{h}(t) + \mathbf{B} \mathbf{x}(t), \\
\mathbf{y}(t) &= \mathbf{C} \mathbf{h}(t),
\end{align}
\end{subequations}
where $\mathbf{h}'(t) \in \mathbb{R}^N$ represents the time derivative of the hidden state $\mathbf{h}(t)$, $\mathbf{A} \in \mathbb{R}^{N \times N}$ is the state transition matrix, $\mathbf{B} \in \mathbb{R}^{N \times D_{\text{in}}}$ and $\mathbf{C} \in \mathbb{R}^{D_{\text{out}} \times N}$ are projection matrices.

To apply continuous-time systems to discrete-time sequences, discretization is necessary. Using the zero-order hold (ZOH)\cite{franklin1998digital} method with a step size $\Delta$, the continuous-time parameters $\mathbf{A}$ and $\mathbf{B}$ are converted into their discrete counterparts $\overline{\mathbf{A}}$ and $\overline{\mathbf{B}}$ as follows:
\begin{subequations}
\label{discretize_param}
\begin{align}
\overline{\mathbf{A}} &= \exp(\Delta \mathbf{A}),\\
\overline{\mathbf{B}} &= (\Delta \mathbf{A})^{-1} (\exp(\Delta \mathbf{A}) - \mathbf{I}) \cdot \Delta \mathbf{B},
\end{align}
\end{subequations}
Combining \eq{eqa_ss} and \eq{discretize_param}, the following discrete-time state space model can be obtained:
\begin{subequations}
\label{discretize_form}
\begin{align}
\mathbf{h}_t &= \bar{\mathbf{A}} \mathbf{h}_{t-1} + \bar{\mathbf{B}} \mathbf{x}_{t-1},\\
\mathbf{y}_t &= \mathbf{C} \mathbf{h}_t,
\end{align}
\end{subequations}

The S4 model efficiently transforms the system from continuous-time parameters $(\Delta, \mathbf{A}, \mathbf{B}, \mathbf{C})$ into discrete-time parameters $(\overline{\mathbf{A}}, \overline{\mathbf{B}}, \mathbf{C})$ by using ZOH and recurrence technology. Usually, the state matrix A is initialized using the HiPPO framework \cite{hippo}, which introduces additional structured components to the state transition matrix $\mathbf{A}$ to effectively enhance the ability of modeling long-term dependencies.

\subsection{Selective SSM}

Building on S4, Mamba \cite{mamba} introduces a data-dependent selection mechanism (illustrated in Fig.~\ref{s6_fig}), giving rise to the selective SSM structure known as S6. In contrast to classical SSMs that rely on static parameterization, S6 dynamically adapts its parameters based on the input, thereby achieving selective attention over each element in the sequence. More formally, given an input sequence $\mathbf{x}$, the model adjusts its transition dynamics in a content-aware manner, i.e., the parameters $\mathbf{B}$, $\mathbf{C}$, and interval $\Delta$ are dynamically projected from the input sequence $\mathbf{x}$ as follows:
\begin{subequations}
\begin{align}
\mathbf{B} &= \text{Linear}_N(\mathbf{x}) \\
\mathbf{C} &= \text{Linear}_N(\mathbf{x}) \\
\Delta &= \text{Softplus}(\text{Parameter}+\text{Projection}_\Delta(\mathbf{x}))
\end{align}
\end{subequations}
where $\text{Linear}_N(\cdot)$ represents projection operation. The operation $\text{Projection}_\Delta(x) = \text{Broadcast}_D(\text{Linear}_1(x))$, first converts $\mathbf{x}$ to one dimension and then expands it to the dimension of $\Delta$ through repeating. This selection mechanism enables Mamba to effectively filter irrelevant noise in temporal tasks while selectively preserving or discarding information related to current input. Additionally, Mamba introduces a hardware-aware algorithm that performs recursive computation of the model through scanning, ensuring effectiveness and efficiency while capturing global contextual information.

\section{Proposed Method}
\label{method}

\begin{figure*}[ht]
\centering
\includegraphics[width=7in]{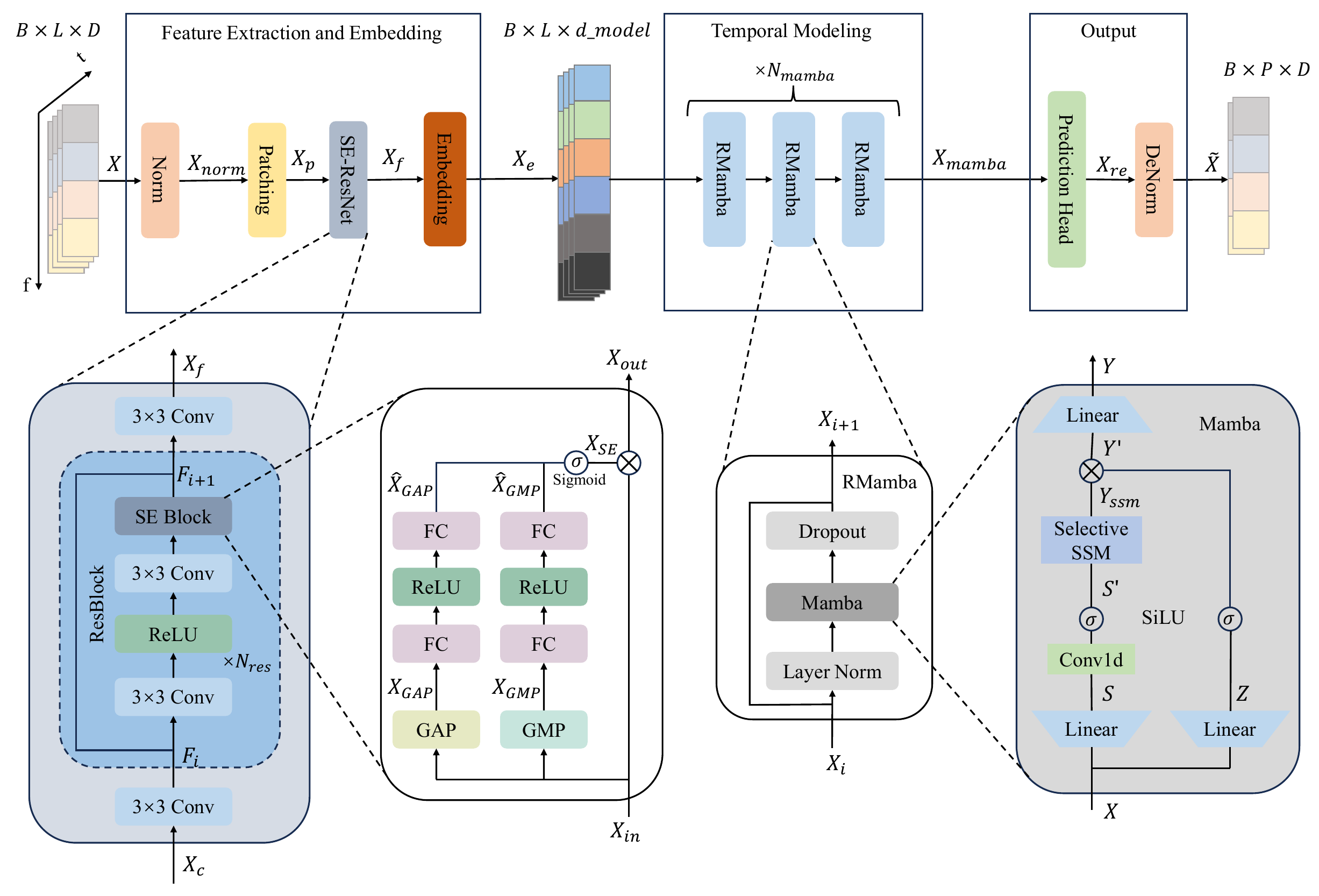}
\caption{Architecture of the CPMamba model. }
\label{arch_abcd}
\end{figure*}
This paper proposes a Mamba-based channel prediction model CPMamba, which consists of three core components: feature extraction and embedding, temporal modeling, and prediction output. The overall architecture is shown in Fig.~\ref{arch_abcd}.

\subsection{Tensor Representation of CSI}
In this paper, CSI prediction is performed in batches. The input data tensor which is a concatenation of the $\mathbf{S}_p\in \mathbb{C}^ {L\times N_t\times K}$ defined in Section \ref{system_model} can be represented as:
\begin{equation}
\mathbf{S}_{\text{batch}} = [\mathbf{S}_p^i]_{i=1}^{B'} \in \mathbb{C}^{B' \times L \times N_t \times K},
\end{equation}
where $B'$ denotes the batch size.
Following the strategy in \cite{llm4cp}, our model processes the CSI of each antenna independently. To this end, the antenna dimension $N_t$ is merged into the batch dimension, resulting in the reshaped tensor:
\begin{equation}
\mathbf{S}'_{\text{batch}} = \text{Rearrange}(\mathbf{S}_{\text{batch}}) \in \mathbb{C}^{B \times L \times K},
\end{equation}
where $B = B' \times N_t$. This restructuring allows the network to handle multi-antenna CSI in a unified and efficient manner. Subsequently, we map the data to the real field by concatenating the real and imaginary parts of the complex features along the feature dimension, yielding the model input $\mathbf{X} \in \mathbb{R}^{B \times L \times D}$, where $D=2K$. Similarly, the predicted output is represented as $\tilde{\mathbf{X}} \in \mathbb{R}^{B \times P \times D}$.

\subsection{Feature Extraction and Embedding Module}
To enhance training stability and mitigate the impact of distribution shift, the input data first undergoes global standardization:
\begin{equation}
\mathbf{X}_{norm}=\frac{\mathbf{X}-\mu}{\sigma},
\end{equation}
where $\mu$ and $\sigma$ are the global mean and standard deviation computed over the entire input tensor, respectively.

Subsequently, to strengthen temporal dependencies across the sequence, a patching operation followed by a linear projection is applied. Following the method in \cite{patch}, the normalized sequence $\mathbf{X}_{norm}\in \mathbb{R}^{B \times L \times D}$ is partitioned into non-overlapping patches of size $N_p$, yielding $\mathbf{X}_{p}\in \mathbb{R}^{B \times {L'} \times D \times N_p}$, where $L'=\lceil\frac{L}{N_p} \rceil$ represents the number of patches. A fully-connected layer is then applied to each patch to perform feature transformation:
\begin{equation}
\mathbf{X}_p=\text{Linear}(\mathbf{X}_{p}).
\end{equation}
In order to transform the real‑imaginary representation of CSI data into a feature space that is more conducive to modeling, a convolutional neural network based on residual blocks enhanced with squeeze‑and‑excitation (SE)\cite{se_block} channel attention mechanisms—termed SE‑ResNet—is adopted for feature extraction.
Specifically, the previously patched sub‑blocks are rearranged to recover the original shape
$\mathbb{R}^{B \times L \times D}$, while the real and imaginary parts are permuted to the channel dimension, producing the input tensor $\mathbf{X}_c \in \mathbb{R}^{B \times {2} \times{L} \times{K}}$. Subsequently, the SE-ResNet is applied to extract features from $\mathbf{X}_c$.

The overall framework of the SE-ResNet is illustrated in the SE-ResNet block diagram in Fig.~\ref{arch_abcd}. Specifically, the process begins with a $3\times3$ convolution layer to expand the channel dimension to $C$, yielding the initial feature representation $\mathbf{F}_0 \in \mathbb{R}^{B \times C \times L \times K}$ as:
\begin{equation}
\mathbf{F}_0 = \text{Conv2d}_{2 \to C}(\mathbf{X}_c),
\end{equation}
where $\text{Conv2d}_{2 \to C}$ denotes a two‑dimensional (2D) convolution that maps the channel dimension from 2 to $C$.
Subsequently, $N_{res}$ residual blocks equipped with a channel attention mechanism are stacked to progressively refine the high‑dimensional channel features:
\begin{equation}
\mathbf{F}_{i+1} = \text{ResBlock}(\mathbf{F}_i) \in \mathbb{R}^{B \times C \times L \times K}, \quad i=0,\dots,N_{res}-1.
\end{equation}
Finally, a $3 \times 3$ convolution is applied to restore the channel dimension, resulting in the residual channel features $X_{f} \in \mathbb{R}^{B \times 2 \times L \times K}$:
\begin{equation}
    \mathbf{X}_{f} = \text{Conv2d}_{C \to 2}(\mathbf{F}_{N_{res}}),
\end{equation}
where $\text{Conv2d}_{C \to 2}$ represents a 2D convolution that maps the channel dimension from $C$ to 2.

The design of the SE block follows \cite{se_block}, which consists of a squeeze phase and an excitation phase. Let $\mathbf{X}_{in}\in \mathbb{R}^{C\times L \times K}$ and $\mathbf{X}_{out}\in \mathbb{R}^{C\times L \times K}$ represent a single tensor in the input batch and output batch of the SE module, respectively. Then the squeeze phase and excitation phase are performed as follows:


\textbf{Squeeze Phase}: Both global average pooling (GAP) and global max pooling (GMP) operations are used to aggregate spatial information in the channel dimension. For input $\mathbf{X}_{in}\in \mathbb{R}^{C\times L \times K}$, the two pooling operations produce statistics $\mathbf{X}_{\rm GAP}\in \mathbb{R}^{C\times1\times 1}$ and $\mathbf{X}_{\rm GMP}\in \mathbb{R}^{C\times1\times 1}$, as follows:
\begin{subequations}
\begin{align}
\mathbf{X}_{\rm GAP}&=\frac{1}{L \times K}\sum_{j=1}^{L}\sum_{k=1}^{K}\mathbf{X}_{in}[:,j,k] \\
\mathbf{X}_{\rm GMP}&=\max_{j,k}\mathbf{X}_{in}[:,j,k].
\end{align}
\end{subequations}

\textbf{Excitation Phase}: The two channel‑wise statistics from the squeeze phase are separately processed by a two‑layer fully‑connected network with shared parameters. The first fully‑connected layer compresses the channel dimension from $C$ to $C/r$, where $r\ge 1$ is a predefined compression ratio. After applying a ReLU activation, the second fully‑connected layer expands the dimension back to $C$, obtaining the refined feature representations of the two branches. Then the outputs of the two branches are added element-wise and activated through the sigmoid function to obtain the final channel attention weight tensor $\mathbf{X}_{\rm SE}$. The specific process is as follows:
\begin{subequations}
\begin{align}
\hat{\mathbf{X}}_{\rm GAP} &= \text{FC}_2(\text{ReLU}(\text{FC}_1(\mathbf{X}_{\rm GAP})))\\
\hat{\mathbf{X}}_{\rm GMP} &= \text{FC}_2(\text{ReLU}(\text{FC}_1(\mathbf{X}_{\rm GMP}))) \\
\mathbf{X}_{\rm SE} &= \sigma(\hat{\mathbf{X}}_{\rm GAP} + \hat{\mathbf{X}}_{\rm GMP})
\end{align}
\end{subequations}
where $\sigma(\cdot)$ represents the sigmoid function, $FC_1:\mathbb{R}^{C \times L \times K} \rightarrow \mathbb{R}^{C/r  \times L \times K}$ and $FC_2:\mathbb{R}^{C/r  \times L \times K} \rightarrow \mathbb{R}^{C \times L \times K}$ denote the two fully‑connected layers. Finally, the input feature $\mathbf{X}_{in}$ is recalibrated channel‑wise by scaling it with $\mathbf{X}_{\rm SE}$ to obtain the feature after channel attention as:
\begin{align}
\mathbf{X}_{\rm out}[i,:,:]=\mathbf{X}_{\rm SE}[i]\times\mathbf{X}_{in}[i,:,:],\quad i=1,\cdots, C,
\end{align}

Multiple such residual blocks equipped with channel attention are stacked (totaling  $N_{res}$ blocks) to progressively enhance the feature‑extraction capability.

In order to adapt the CNN-processed features into a representation suitable for the Mamba model,  $\mathbf{X}_{f}$ is first reshaped into $\mathbf{X}_{\rm f'}\in \mathbb{R}^{B \times L\times D}$, Subsequently, an Embedding operation consisting of a fully connected layer maps these features to the Mamba model’s designated input dimension, yielding $\mathbf{X}_{\rm e}\in \mathbb{R}^{B\times L \times {d_{model}}}$.

\subsection{Mamba-Based Temporal Modeling Network}
After the data embedding is obtained, we employ multiple residual Mamba (RMamba) modules connected in series to perform temporal modeling. The overall architecture is illustrated in Fig.~\ref{arch_abcd}. Each Mamba block implements the following transformation:
\begin{equation}
\label{mamba_eqa}
\mathbf{X}_{i+1} = \mathbf{X}_{i} + \text{Dropout}(\text{Mamba}(\text{Norm}(\mathbf{X}_{i}))),
 \end{equation}
where $\text{Mamba}(\cdot)$ represents the standard Mamba module and $\text{Norm}(\cdot)$ represents LayerNorm.

The standard Mamba module employs a selective SSM as its core mechanism for temporal modeling. Given an input sequence $\mathbf{X} \in \mathbb{R}^{B \times L \times d_{model}}$, it returns an output sequence $\mathbf{Y} \in \mathbb{R}^{B \times L \times d_{model}}$ of identical shape. The forward pass is structured into two parallel branches—one for state‑space modeling and the other for gating—implemented as follows:

First, two independent linear projections map the input into an expanded feature space:
\begin{subequations}
\begin{align}
\mathbf{S} &= \text{Linear}_S(\mathbf{X}) \in \mathbb{R}^{B \times L \times E}, \\
\mathbf{Z} &= \text{Linear}_Z(\mathbf{X}) \in \mathbb{R}^{B \times L \times E},
\end{align}
\end{subequations}
where $E=d_{model}\times {expand}$ denotes the expanded dimension and $expand$ is the tunable expansion coefficient. The dual-path projection design allows separate processing of temporal dynamic features and gating adjustment signals.

Within the state‑space modeling branch, the feature $\mathbf{S}$ passes through a causal convolution layer to incorporate local temporal dependencies. Causal convolution ensures that the model only depends on historical information. Following this, non‑linear activation is applied via a SiLU activation function:
\begin{equation}
\mathbf{S}' = \text{SiLU}(\text{Conv1d}(\mathbf{S})) \in \mathbb{R}^{B \times L \times E}.
\end{equation}

The processed feature $\mathbf{S}'$ is then fed into the selective state space model for temporal modeling:
\begin{equation}
\mathbf{Y}_{\text{ssm}} = \text{Selective\_SSM}(\mathbf{S}') \in \mathbb{R}^{B \times L \times E}.
\end{equation}

This module dynamically adjusts the state‑transition process through an input‑dependent selection mechanism, thereby capturing complex temporal dependencies in channel sequences while retaining linear computational complexity. (The detailed structure is illustrated in Fig.~\ref{s6_fig} of Section. \ref{preliminary}.)

During the feature‑fusion stage, the gating branch $\mathbf{Z}$ is activated via a SiLU function and then combined with the SSM output through element‑wise multiplication, enhancing the model’s ability to adaptively modulate different time steps and feature dimensions:
\begin{equation}
\label{eqa_gate}
\mathbf{Y}' = \text{SiLU}(\mathbf{Z}) \odot \mathbf{Y}_{\text{ssm}} \in \mathbb{R}^{B \times L \times E}.
\end{equation}

Finally, the fused feature is projected back to the original channel dimension through a linear layer:
\begin{equation}
\mathbf{Y} = \text{Linear}_{\text{out}}(\mathbf{Y}') \in \mathbb{R}^{B \times L \times d_{model}}.
\end{equation}

Multiple RMamba modules are stacked in series over $N_{mamba}$ layers, ultimately yielding the temporally modeled representation $\mathbf{X}_{\rm mamba} \in \mathbb{R}^{B\times L \times {d_{model}}}$.

\subsection{Multi-Step Prediction Output Module}
The prediction head transforms the output of the temporal modeling stage into the predicted future channel coefficients. First, a fully connected layer operating along the feature dimension projects the output of the Mamba stage $X_{mamba}\in \mathbb{R}^{B \times L \times d_{model}}$ into a space whose dimensionality matches the original CSI representation:
\begin{equation}
\mathbf{X}_{fe} = \text{FC}_{f}(\mathbf{X}_{mamba}) \in \mathbb{R}^{B \times L \times D}  \label{fc_eqa1}
\end{equation}
Next, a second fully connected layer—applied along the temporal axis—maps the historical sequence length $L$ to the future prediction horizon $P$. To enable the layer to operate on the temporal dimension, the last two dimensions of $X_{fe}$ are transposed before and after the linear transformation:
\begin{equation}
{\mathbf{X}}_{re} = \text{Permute}(\text{FC}_{t}(\text{Permute}(\mathbf{X}_{fe}))) \in \mathbb{R}^{B \times P \times D} \label{fc_eqa2}
\end{equation}
where $\text{Permute}(\cdot)$ (swaps  the  feature  and  temporal  dimensions. Finally, the normalized predictions are reverted to the original CSI scale using the statistics preserved from the input normalization stage:
\begin{equation}
\label{denorm_eqa}
\tilde{\mathbf{X}} = \sigma {\mathbf{X}}_{re} + \mu \in \mathbb{R}^{B\times P \times D}
\end{equation}
Here, $\mu$ and $\sigma$ denote the global mean and standard deviation computed during the input normalization phase.

\subsection{The CPMamba Prediction Algorithm}
The Mamba-based CSI prediction algorithm is given in Algorithm \ref{alg:csi_mamba}, with specific description as follows:

\textbf{Data preprocessing} (lines 3-5): First, the raw input $\mathbf{X}$ is normalized by its global statistics to produce $\mathbf{X}_{norm}$, reducing distribution shift.
Next, $\mathbf{X}_{norm}$ is partitioned along the time dimension into non-overlapping sub‑blocks of size $N_p$. Each sub‑block is then linearly projected to obtain the final block‑level representation $\mathbf{X}_{p}$.

\textbf{SE-ResNet feature extraction} (lines 6-13): The block features $\mathbf{X}_p$ are reshaped into $\mathbf{X}_c$, suitable for convolutional processing. A $3 \times 3$ convolution first expands the channel dimension to $C$. The result is then passed through $N_{res}$ residual blocks equipped with the SE channel attention mechanism to extract high‑level features $\mathbf{X}_{f}$. After rearrangement, $\mathbf{X}_{f'}$ is mapped via an embedding layer to the dimension required by the Mamba backbone, producing the embedded representation $\mathbf{X}_{e}$.

\textbf{Temporal modeling network} (lines 14-18): Taking $\mathbf{X}_{e}$ as input, the data flows sequentially through $N_{mamba}$ RMamba modules. Each module sequentially performs layer normalization, selective SSM, Dropout, and residual connection, progressively refining the temporal representation. The output $\mathbf{X}_{\rm mamba}$ captures long‑range dependencies across the sequence.

\textbf{Multi-step prediction output} (lines 19-21): Finally, $\mathbf{X}_{\rm mamba}$ is projected back to the original CSI feature space via fully‑connected layer $\text{FC}_f$, yielding $\mathbf{X}_{fe}$. Then, a second layer $\text{FC}_t$ maps the historical length to future prediction horizon $P$, resulting in $\mathbf{X}_{re}$. A denormalization step is applied to restore the physical scale, producing the final CSI prediction $\tilde{\mathbf{X}}$.

\begin{algorithm}[ht]
\caption{CPMamba Prediction Algorithm}
\label{alg:csi_mamba}
\begin{algorithmic}[1]
\STATE \textbf{Input:} Tensor representation of historical CSI sequence $\mathbf{X} \in \mathbb{R}^{B\times L\times D}$
\STATE \textbf{Output:} Tensor representation of predicted future CSI sequence $\tilde{\mathbf{X}} \in \mathbb{R}^{B\times P\times D}$

\STATE \textbf{$\triangleright$ Data Preprocessing}
\STATE Normalization: $\mathbf{X}_{\rm norm} = (\mathbf{X}-\mu)/\sigma$
\STATE Patching along time dimension and linear projection: $\mathbf{X}_p = \text{Linear}(\mathbf{X}_p)$

\STATE \textbf{$\triangleright$ SE-ResNet Feature Extraction}
\STATE Reshape to $\mathbf{X}_c \in \mathbb{R}^{B \times 2 \times L \times K}$
\STATE $3\times3$ convolution to expand channels to $C$
\FOR{$i = 1$ to $N_{res}$}
    \STATE SE-ResNet feature extraction
\ENDFOR
\STATE $3\times3$ convolution to restore channel dimension, output $\mathbf{X}_{f}$
\STATE Reshape to $\mathbf{X}_{\rm f'} \in \mathbb{R}^{B\times L\times D}$, embed with fully connected layer to obtain $\mathbf{X}_{\rm e} \in \mathbb{R}^{B\times L\times d_{model}}$

\STATE \textbf{$\triangleright$ Temporal Modeling Network}
\FOR{$i = 1$ to $N_{mamba}$}
    \STATE Residual Mamba sequence modeling: \eq{mamba_eqa}
\ENDFOR
\STATE Obtain temporal features $\mathbf{X}_{\text{mamba}}$

\STATE \textbf{$\triangleright$ Multi-Step Prediction Output}
\STATE Two-layer fully connected transformation for feature space and temporal prediction steps, as in \eq{fc_eqa1}-\eqref{fc_eqa2}.
\STATE Denormalization to restore CSI, output $\tilde{\mathbf{X}}$, as in \eq{denorm_eqa}.
\end{algorithmic}
\end{algorithm}

\subsection{Complexity Analysis}
This section presents a computational complexity analysis of the proposed model, with particular emphasis on the number of multiplication operations required during online inference.
The computational complexity of the proposed model is primarily attributed to several key modules:
\begin{itemize}
\item Data preprocessing incurs a complexity of $O(B \times \frac{L}{N_p} \times D \times N_p^2)$, where the patch size is set to $N_p=4$ and $D=2K$ is the input dimension.
\item Residual convolutional network consists of $N_{res}=4$ residual blocks, each comprising a 3×3 convolutional layer and a channel attention mechanism. Its complexity amounts to $O(N_{res} \times B \times L \times K \times C_{in} \times C_{out} \times 3^2)$, with $C_{in}=64$ and $C_{out}=64$.
\item Input embedding layer projects the input from dimension $D$ to $d_{model}=768$, contributing a complexity of $O(B \times L \times D \times d_{model})$.
\item Mamba backbone module incorporates $N_{mamba}$ RMamba blocks, whose complexity is dominated by the selective state-space mechanism and scales as $O(N_{mamba} \times B \times L \times d_{model} \times expand \times (d_{conv} + d_{state}))$, where $expand=2$, $d_{conv}=4$, and $d_{state}=4$.
\item Dimension projection layer maps features from $d_{model}$ back to output dimension $D$, with complexity $O(B \times L \times d_{model} \times D)$; Temporal prediction head further maps the sequence from length $L=16$ to $P=4$, adding a complexity $O(B \times D \times L \times P)$.
\end{itemize}

It can be found that the main complexity of the proposed model comes from the temporal complexity brought by the residual convolutional network. For the Mamba modules of the temporal modeling model, its complexity is nearly linear. Compared to the standard Transformer architecture—whose complexity is characterized by the quadratic self-attention operation $O(B \times L^2 \times d_{model})$ and the feed-forward network operation $O(B \times L \times d_{model}^2)$---the present model demonstrates superior scalability for longer sequences. This advantage is achieved by utilizing a selective state-space model as the core mechanism for time-series modeling, which maintains linear complexity in sequence length, while confining quadratic-complexity operations solely to localized convolutions and linear projections.

\section{Experiments}
\label{experiment}
This chapter details the experimental setup and training parameters, and demonstrates the effectiveness of the proposed method through comprehensive comparisons with baseline methods and ablation studies.

\subsection{Experimental Setup}
\subsubsection{Dataset Configuration}
The channel dataset utilized in this experiment is generated using the Quasi Deterministic Radio Channel Generator (QuaDRiGa) \cite{quadriga}, strictly adhering to 3GPP standards \cite{3gpp}. The system operates at a center frequency of $f_c = 2.4$ GHz. The base station is equipped with a $4\times4$ dual-polarized UPA (total antennas $N_t=32$) with an element spacing of $d = \lambda_c / 2$, where $\lambda_c$ denotes the wavelength corresponding to $f_c$, while the UE is equipped with a single omnidirectional antenna. The simulation generates a total bandwidth of 17.28 MHz ($2K=96$ subcarriers), which is partitioned into two adjacent bands: the lower $K=48$ subcarriers (8.64 MHz) are allocated to the uplink, and the upper $K=48$ subcarriers are allocated to the downlink.

In terms of spatial configuration, the base station is fixed, while UEs move along linear trajectories. UEs are randomly initialized within a 20 m to 50 m radius relative to a cluster center located 200 m from the base station. Regarding the propagation environment, the Urban Macro-cell (UMa) NLoS scenario is adopted for model training and primary testing, whereas the Urban Micro-cell (UMi) LoS/NLoS scenarios are employed to evaluate cross-scenario generalization capability.

The training and validation datasets contain 8,000 and 1,000 samples, respectively, with user speeds uniformly distributed across $[10, 100]$ km/h. Separately, the test dataset is generated to evaluate 10 discrete velocities equally spaced within the same range, comprising 1,000 samples per velocity. Structurally, each sample consists of $L=16$ historical frames and $P=4$ future frames for both uplink and downlink channels, captured at a sampling interval of 0.5 ms.

\subsubsection{Baselines}
To verify the effectiveness of the proposed CPMamba model, the following representative channel prediction benchmark models are used for comparison:
\begin{itemize}
\item \textbf{No Prediction (NP)}: Directly uses the last frame of the historical channel sequence as the prediction result, serving as a baseline to gauge the necessity and performance of predictive models.
\item \textbf{PAD \cite{pad}}: The Prony-based Angle-Delay (PAD) method exploits the sparsity of physical models, estimating channel angle-delay characteristics through signal parameter estimation to predict future channels. It represents a classic model-driven approach.
\item \textbf{LSTM \cite{lstm}}: A Long Short-Term Memory network that effectively captures temporal correlations in channel sequences through gating mechanisms.
\item \textbf{CNN \cite{cnn_cp}}: A Convolutional Neural Network-based prediction algorithm that treats time-frequency 2D CSI data as images. It employs a convolutional autoencoder structure to extract local spatiotemporal features, followed by a convolutional layer to map feature representations to prediction results.
\item \textbf{Transformer \cite{transformer}}: A Transformer-based channel prediction method utilizing self-attention mechanisms to simultaneously model historical information and predict future channels, thereby mitigating sequential error propagation.
\item \textbf{LLM4CP \cite{llm4cp}}: An LLM-based channel prediction scheme achieving cross-modal knowledge transfer by fine-tuning a pre-trained GPT-2 model. It represents a prediction model with strong generalization capabilities.
\item \textbf{ChannelMamba \cite{channelmamba}}: The first model to apply Mamba to channel prediction, utilizing dual-domain input fusion (frequency-domain CSI and delay-domain CIR) and cross-path parameter sharing strategies to efficiently capture temporal dynamic dependencies.
\end{itemize}

To ensure a fair comparison, all deep learning-based baseline algorithms process antenna dimension data in parallel under identical hardware and environmental conditions, using Normalized Mean Square Error (NMSE) as the training loss function. Note that PAD is a non-learning method and is applied exclusively in TDD systems.

\subsubsection{Performance Metrics}
Model performance is evaluated using the following three standard metrics:
\begin{itemize}
\item \textbf{NMSE}: Defined in \eq{nmse_eqa}, this metric is widely used in CSI-related tasks as the primary objective. It evaluates the normalized error energy between the predicted and ground truth CSI.
\item \textbf{RMSE (Root Mean Square Error)}: Measures the square root of the average squared deviation. Due to the squaring operation, RMSE is highly sensitive to outliers, assigning heavier penalties to larger errors, given by:
\begin{equation}
\label{rmse_eqa}
\text{RMSE}(\hat{\mathbf{S}}_{f},\mathbf{S}_{f}) = \sqrt{\mathbb{E}\left[ |\hat{\mathbf{S}}_{f}-\mathbf{S}_{f}|^2 \right]}
\end{equation}
where $\mathbb{E}[\cdot]$ denotes the mean operation over all elements in the tensor.

\item \textbf{MAE (Mean Absolute Error)}: Measures the average absolute error between predicted and true values. Unlike RMSE, it treats errors linearly and is less sensitive to outliers, providing a direct and robust reflection of the average error magnitude, as follows:
\begin{equation}
\label{mae_eqa}
\text{MAE}(\hat{\mathbf{S}}_{f},\mathbf{S}_{f}) = \mathbb{E}\left[ |\hat{\mathbf{S}}_{f}-\mathbf{S}_{f}| \right]
\end{equation}
\end{itemize}

Given the significant dynamic range of CSI data, NMSE is used as the primary evaluation metric, with RMSE and MAE serving as auxiliary metrics.

\subsubsection{Training Configuration}
To enhance model generalization and robustness, random noise (AWGN) with an SNR uniformly sampled from $[5, 20]$ dB is injected into the training data. The training process spans a total of 300 epochs. We employ a two-stage learning rate schedule: the learning rate is initialized at $lr_{start}$ for the first $N_{start}$ epochs to ensure fast convergence, and then decayed to $lr_{end}$ for the subsequent $N_{end}$ epochs to facilitate fine-tuning. The optimizer and specific hyperparameters for both the training strategy and the network architecture are detailed in Table \ref{params_settings}.
\begin{table}[t]
	\small
	\caption{Hyper-parameter Settings}
	\label{params_settings}
	\centering
	\begin{tabular}{lc}
		\toprule
		\textbf{Parameter} & \textbf{Value} \\
        \midrule
        \multicolumn{2}{l}{\textit{Training Settings}} \\
		Batch Size  & 256 \\
		Total Epochs & 300 \\
		Optimizer & Adam ($\beta_1=0.9, \beta_2=0.999$) \\
		Initial Learning Rate ($lr_{start}$) & $1 \times 10^{-3}$ \\
        Initial Training Epochs ($N_{start}$) & 200 \\
        Fine-tuning Learning Rate ($lr_{end}$) & $1 \times 10^{-4}$ \\
        Fine-tuning Epochs ($N_{end}$) & 100 \\
        \midrule
        \multicolumn{2}{l}{\textit{CPMamba Architecture}} \\
        Residual Layers ($N_{res}$) & 4 \\
        Mamba Layers ($N_{mamba}$) & 6 \\
        Model Dimension ($d_{model}$) & 768 \\
        SSM State Dimension ($d_{state}$) & 4 \\
        Conv Kernel Size ($d_{conv}$) & 4 \\
        Expansion Factor ($expand$) & 2 \\
        \bottomrule
	\end{tabular}
\end{table}

\subsection{Performance Evaluation}
To comprehensively assess model performance, we conduct comparative experiments across varying user mobility speeds and SNR conditions. All models are trained on the UMa-NLoS scenario.

\subsubsection{Impact of User Mobility Speed}
We evaluate the robustness of different models at 10 discrete velocities equally spaced within the range of $[10, 100]$ km/h.

\begin{figure}[t]
\centering
\includegraphics[width=7.5cm]{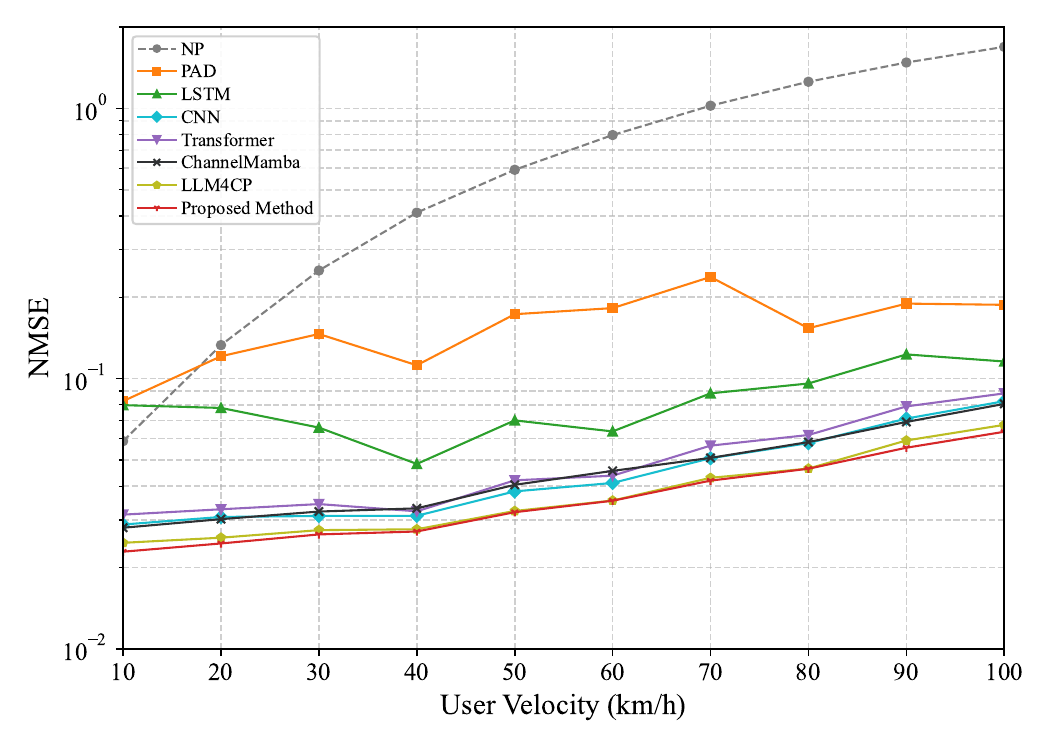}
\caption{NMSE performance versus user mobility speed in TDD systems.}
 \label{fig_tdd_uv}
\end{figure}

Fig.~\ref{fig_tdd_uv} illustrates the NMSE performance in the TDD system. As the user speed increases, the NMSE of all models generally exhibits an upward trend. This confirms that high mobility accelerates channel aging, reducing the temporal correlation and increasing prediction difficulty. The "No Prediction" baseline deteriorates rapidly, with NMSE exceeding 1.0 at 70 km/h, highlighting the necessity of prediction algorithms. In contrast, the proposed CPMamba maintains the lowest NMSE across the entire speed range (below 0.07 even at 100 km/h). Notably, it achieves performance comparable to the heavy-weight LLM4CP while operating with significantly lower computational complexity, validating the efficiency of the Mamba architecture in capturing temporal dynamics.

\begin{figure}[t]
\centering
\includegraphics[width=7.5cm]{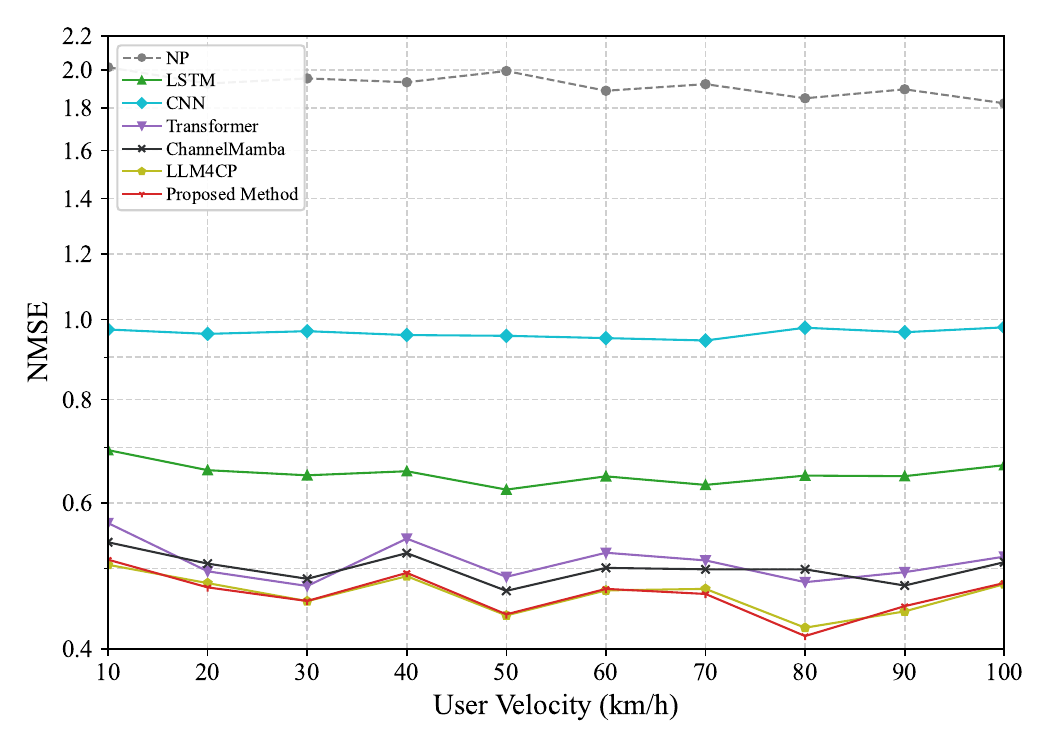}
\caption{NMSE performance versus user mobility speed in FDD systems.}
 \label{fig_fdd_uv}
\end{figure}

Fig.~\ref{fig_fdd_uv} presents the results for the FDD system. Unlike the TDD scenario, the performance in FDD (predicting downlink CSI from historical uplink CSI) shows weak correlation with user speed. This suggests that the dominant error source in FDD is the difficulty of modeling the complex non-linear mapping between heterogeneous frequency bands, rather than mobility-induced channel aging. While spatial-feature-based models like CNN struggle to bridge this frequency gap (NMSE $\approx$ 1.0), CPMamba demonstrates superior capability in modeling cross-band dependencies, maintaining an NMSE generally below 0.6.


\subsubsection{Impact of SNR Variations}
To evaluate noise robustness, we fix the user speed at 60 km/h and vary the SNR from 0 dB to 25 dB.

\begin{figure}[t]
\centering
\includegraphics[width=7.5cm]{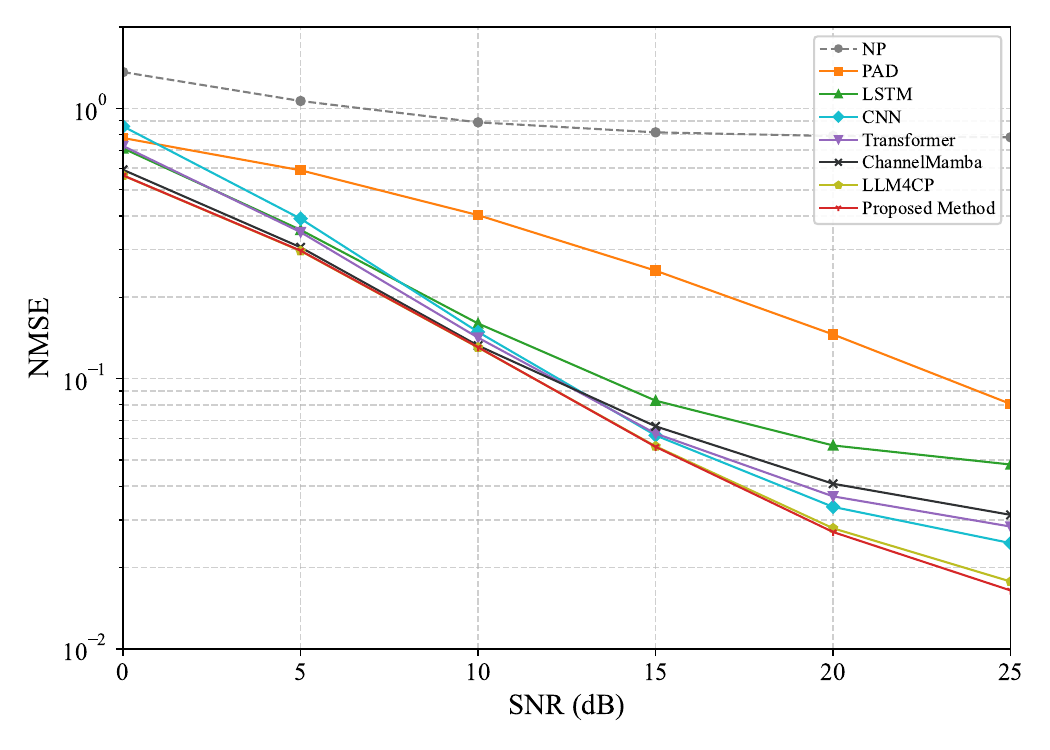}
\caption{NMSE performance versus SNR in TDD systems.}
 \label{fig_tdd_snr}
\end{figure}

\begin{figure}[t]
\centering
\includegraphics[width=7.5cm]{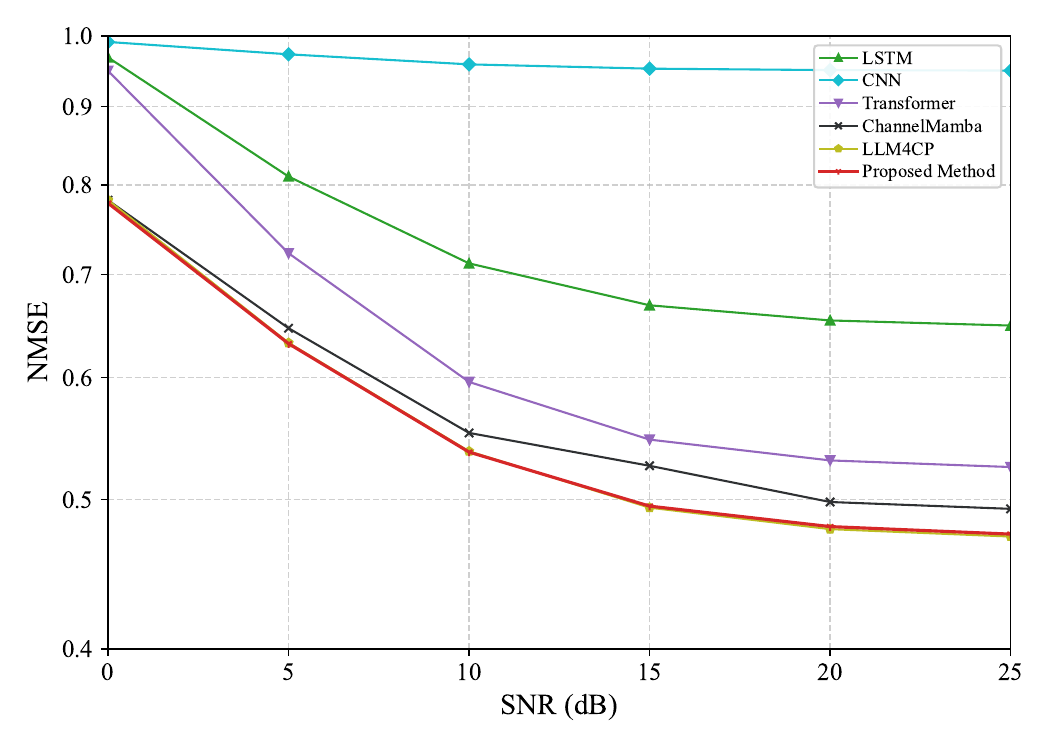}
\caption{NMSE performance versus SNR in FDD systems.}
 \label{fig_fdd_snr}
\end{figure}

Fig.~\ref{fig_tdd_snr} and Fig.~\ref{fig_fdd_snr} depict the results for TDD and FDD systems, respectively. As expected, prediction accuracy improves for all models as SNR increases. CPMamba consistently exhibits superior noise robustness in both systems. In the TDD system, it achieves an NMSE below 0.02 at 25 dB, significantly outperforming LSTM and Transformer. In the FDD system, CPMamba maintains an NMSE below 0.8 even at 0 dB. These results verify that CPMamba matches the robustness of large-scale pre-trained models (LLM4CP) without relying on massive pre-training data.


\subsubsection{Evaluation using RMSE and MAE Metrics}
Since NMSE serves as the training objective, we introduce RMSE and MAE as auxiliary metrics to provide a more unbiased evaluation. The average metric values across all tested speeds are summarized in Table \ref{table_rmse_mae_tdd} and Table \ref{table_rmse_mae_fdd}.

In the TDD system, CPMamba achieves the best performance in both RMSE and MAE, slightly outperforming LLM4CP. In the FDD system, where errors are generally higher due to weak spectral correlations, CPMamba achieves an RMSE of 0.3341 and MAE of 0.2064. While this performance is comparable to that of LLM4CP (RMSE 0.3342), it is crucial to note that CPMamba achieves this with a significantly smaller parameter footprint and lower inference latency. This demonstrates that CPMamba effectively strikes an optimal balance between prediction accuracy and system efficiency.

\begin{table*}[ht]
\small
\caption{Average RMSE and MAE performance in TDD systems.}
\label{table_rmse_mae_tdd}
\centering
\begin{tabular}{lccccccc}
\toprule
Metric & NP & PAD & LSTM & CNN & Transformer &  LLM4CP & CPMamba \\
\midrule
RMSE  & 0.3984 & 0.0983 & 0.1395 & 0.1037 & 0.1077 & 0.0948 & \textbf{0.0932} \\
MAE & 0.2540 & 0.0774 & 0.0979 & 0.0771 & 0.0800 & 0.0725 & \textbf{0.0716} \\
\bottomrule
\end{tabular}
\end{table*}

\begin{table*}[ht]
\small
\caption{Average RMSE and MAE performance in FDD systems.}
\label{table_rmse_mae_fdd}
\centering
\begin{tabular}{lcccccc}
\toprule
Metric & NP & LSTM & CNN & Transformer &  LLM4CP & CPMamba \\
\midrule
RMSE  & 0.6810 & 0.3963 & 0.4828 & 0.3499 & 0.3342 & \textbf{0.3341} \\
MAE & 0.4477 & 0.2552 & 0.3223 & 0.2198 & 0.2110 & \textbf{0.2064} \\
\bottomrule
\end{tabular}
\end{table*}

\subsection{Generalization Experiments}

\begin{figure}[t]
\center{\includegraphics[width=7.5cm]  {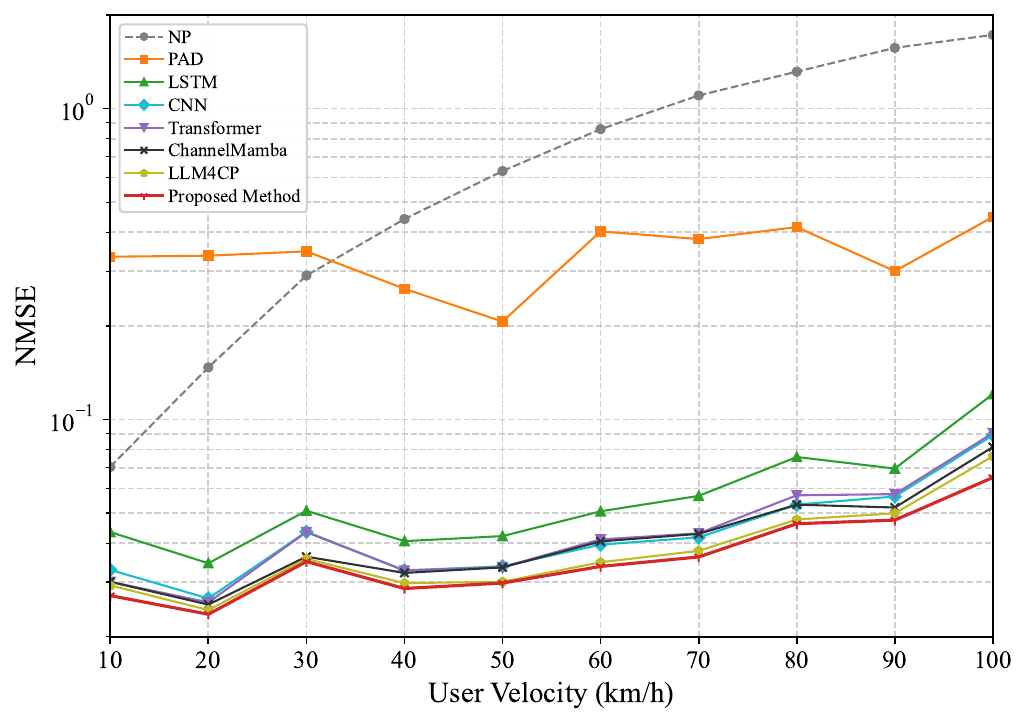}}
\caption{NMSE performance under different user mobility speeds in UMi-NLOS scenario for TDD system.}
 \label{tdd_uv_Umi_NLOS}
\end{figure}
\begin{figure}[t]
\center{\includegraphics[width=7.5cm]  {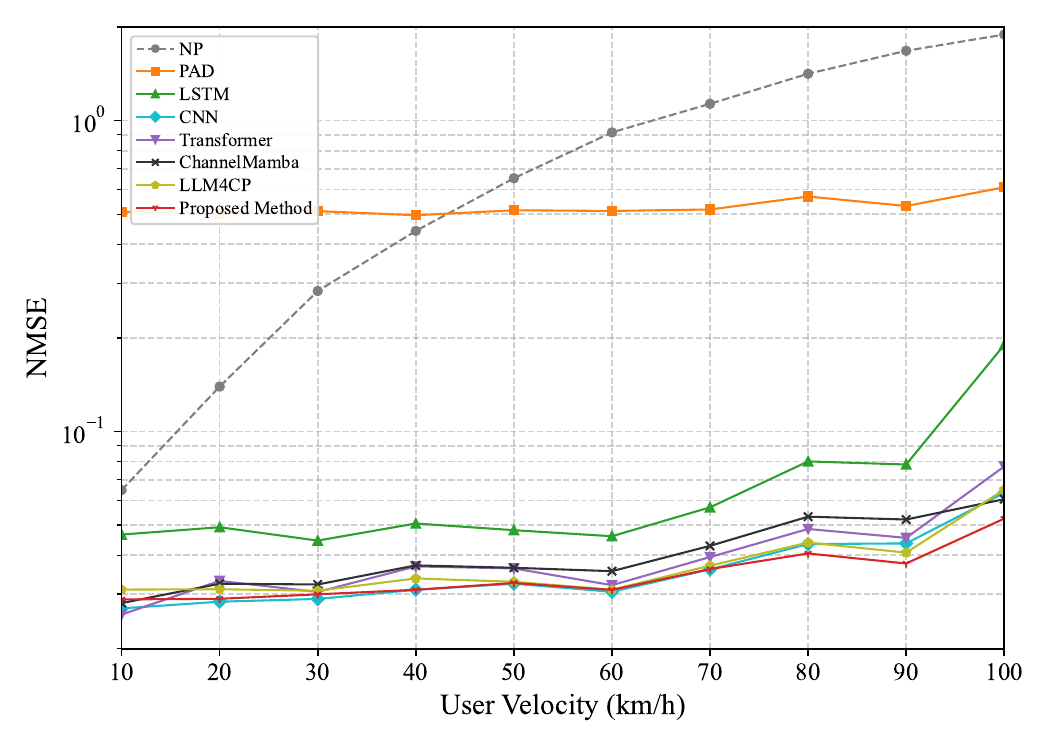}}
\caption{NMSE performance under different user mobility speeds in UMi-LOS scenario for TDD system.}
 \label{tdd_uv_Umi_LOS}
\end{figure}

To evaluate the cross-scenario generalization capability, models trained exclusively on the UMa-NLoS dataset are directly evaluated on the UMi-NLoS and UMi-LOS scenarios without any fine-tuning.

Fig.~\ref{tdd_uv_Umi_NLOS} and Fig.~\ref{tdd_uv_Umi_LOS} illustrate the performance under this challenging domain-shift setting. Compared to the native UMa scenario (Fig.~\ref{fig_tdd_uv}), the NMSE for all models inevitably increases due to the significant statistical discrepancies in channel characteristics between UMi-NLoS and UMi-LOS scenarios.

Despite these challenges, CPMamba consistently maintains the lowest NMSE in both UMi-NLoS and UMi-LOS scenarios. This superior generalization is attributed to the Selection Mechanism inherent in the Mamba block. Unlike varying LSTM or Transformer structures with fixed weights, the input-dependent SSM parameters allow CPMamba to act as a content-aware filter, dynamically adjusting its state transition dynamics based on the current input context. This enables the model to effectively extract robust temporal features that are invariant across different propagation environments.

\begin{table*}[ht]
\small
\centering
\caption{Results of Ablation Study}
\label{ablation_table}
\begin{tabular}{lcccc}
\toprule
Metric & CPMamba & w/o SE-ResNet & w/o patching & Replace w/ Trans. \\
\midrule
NMSE & \textbf{0.0375} & 0.0666 & 0.0381 & 0.0450 \\
RMSE & \textbf{0.0931} & 0.1237 & 0.0938 & 0.0962 \\
MAE & \textbf{0.0715} & 0.0909 & 0.0719 & 0.0737 \\
\bottomrule
\end{tabular}
\end{table*}

\begin{table*}[ht]
\small
\centering
\caption{Computational Efficiency Comparison}
\label{efficiency_table}
\begin{tabular}{lccccc}
\toprule
\textbf{Metric} & \textbf{LSTM} & \textbf{CNN} & \textbf{Transformer} & \textbf{LLM4CP} & \textbf{CPMamba} \\
\midrule
Parameters (M) & 1.140 & 3.147 & 1.765 & 0.976 & \textbf{0.487} \\
Inference Time (ms) & 17.785 & 13.854 & \textbf{11.255} & 23.107 & 15.801 \\
\bottomrule
\end{tabular}
\end{table*}

\subsection{Ablation Study}

To rigorously evaluate the contribution of each core component within CPMamba, we conducted ablation studies by systematically modifying the model architecture. The experiments were performed in the TDD system, with metrics averaged across 10 distinct user speeds ranging from 10 km/h to 100 km/h. The results are summarized in Table \ref{ablation_table}, where three variants are compared against the full model:

\begin{itemize}
\item \textbf{w/o CNN}: The removal of SE-ResNet leads to the most severe performance degradation, with NMSE increasing from 0.0375 to 0.0666 (an approx. 77\% error rise). This indicates that the Mamba backbone relies heavily on high-quality feature embeddings. The CNN is essential for filtering noise and extracting spatial-spectral patterns, enabling Mamba to focus on modeling temporal dependencies.

\item \textbf{w/o Patching}: Although removing patching results in a relatively minor NMSE increase (from 0.0375 to 0.0381), this module is computationally lightweight. It remains a cost-effective component for aggregating local temporal context and enhancing input representation without imposing significant overhead.

\item \textbf{w/ Transformer}: Replacing the Mamba backbone with a standard Transformer results in a noticeable accuracy drop, with NMSE rising to 0.0450 and RMSE to 0.0962. This confirms that the selective state-space mechanism in CPMamba is more suitable than self-attention for capturing the continuous dynamics of channel evolution.
\end{itemize}

Overall, the complete CPMamba architecture achieves the best performance across all metrics, confirming that the synergistic integration of SE-ResNet for feature extraction and Mamba for temporal modeling is essential for accurate channel prediction.

\subsection{Efficiency Comparison}
Table \ref{efficiency_table} provides a detailed evaluation of computational efficiency, focusing on model size and actual inference latency. These metrics are critical for real-time implementation in BS, where the system must handle massive user access with strict latency constraints.

In terms of parameter efficiency, CPMamba achieves the most lightweight architecture among all baselines, requiring only 0.487 M parameters. This represents a remarkable reduction of approximately 50\% compared to the large-scale LLM4CP (0.976 M) and over 84\% compared to the CNN (3.147 M). Even compared to standard Transformer and LSTM models, CPMamba utilizes significantly fewer parameters. Such compactness drastically alleviates the memory overhead at the BS, enabling it to efficiently maintain and execute prediction models for a large number of concurrent users.

Regarding inference speed, the Transformer performs best (11.255 ms) due to its highly parallelizable attention mechanism. However, CPMamba (15.801 ms) remains highly competitive, delivering an inference speed that is approximately 31.6\% faster than the state-of-the-art LLM4CP (23.107 ms). While the CNN achieves relatively fast inference, its massive parameter count creates a memory bottleneck, limiting the scalability of the BS when serving dense user populations.

Overall, CPMamba strikes an optimal balance between model size and execution speed. It matches the high-accuracy capabilities of complex models like LLM4CP while maintaining a significantly lighter footprint and faster response time, validating the efficiency of the selective state space model for real-time, high-throughput channel prediction tasks at the network edge.

\section{Conclusion}
\label{conclusion}
In this paper, we proposed CPMamba, a channel prediction framework based on selective state space models, designed for high-mobility MIMO-OFDM systems. This framework accurately maps historical uplink CSI to future downlink CSI. To ensure effective feature representation, we designed a feature extraction module integrating patching operations and an SE-ResNet, transforming raw CSI into robust high-dimensional embeddings. For temporal modeling, the framework employs a backbone of stacked residual Mamba modules. By leveraging the selective state space mechanism, it effectively captures long-range dependencies in channel sequences while maintaining linear computational complexity. Extensive experimental results demonstrate that CPMamba achieves superior performance in both TDD and FDD systems. It exhibits strong robustness against noise and excellent generalization capabilities across varying user speeds.

Future work will focus on two main aspects: exploring more efficient data preprocessing techniques to further enhance model performance, and evaluating the model with real-world channel measurement data to verify its practical applicability.

\bibliographystyle{IEEEtran}
\bibliography{refs}








\end{document}